# Superconductivity of $Ca_2InN$ with a layered structure embedding anionic indium chain array

Sehoon Jeong[1], Satoru Matsuishi[2], Kimoon Lee[3], Yoshitake Toda[3], Sung Wng Kim[4,5] and Hideo Hosono[1,2,3,*]

[1]Materials and Structures Laboratory, Tokyo Institute of Technology, 4259 Nagatsuta-cho, Midori-ku, Yokohama, Japan

[2]Materials Research Center for Element Strategy, Tokyo Institute of Technology, 4259 Nagatsuta-cho, Midori-ku, Yokohama, Japan

[3]Frontier Research Center, Tokyo Institute of Technology, 4259 Nagatsuta-cho, Midori-ku, Yokohama, Japan

[4]Department of Energy Science, Department of Physics, Sungkyunkwan University, Suwon, Korea.

[5]IBS Center for Integrated Nanostructure Physics, Institute for Basic Science, Sungkyunkwan University, Suwon, Korea.

E-mail: hosono@msl.titech.ac.jp

**Abstract**

We report the emergence of superconductivity in $Ca_2InN$ consisting of a 2-dimensional (2D) array of zigzag indium chains embedded between $Ca_2N$ layers. A sudden drop of resistivity and a specific-heat ($C_p$) jump attributed to the superconducting transition were observed at 0.6 K. The Sommerfeld coefficient $\gamma$ = 4.24 mJ/mol K$^2$ and Debye temperature $\Theta_D$ = 322 K were determined from the $C_p$ of the normal conducting state and the superconducting volume fraction was estimated to be ~80% from the $C_p$ jump, assuming a BCS-type weak coupling. Density functional theory calculations demonstrated that the electronic bands near the Fermi level ($E_F$) are mainly derived from In 5$p$ orbitals with $\pi$ and $\sigma$ bonding states and the Fermi surface is composed of cylindrical parts, corresponding to the quasi-2D electronic state of the In chain array. By integrating the projected density of states of In-$p$ component up to $E_F$, a valence electron population of ~1.6 electrons/In was calculated, indicating that the partially anionic state of In. The In 3$d$ binding energies observed in $Ca_2InN$ by x-ray photoemission spectroscopy were negatively shifted from that in In metal. The superconductivity of $Ca_2InN$ is associated with the $p$-$p$ bonding states of the anionic In layer.

## 1. Introduction

Superconductivity of magnesium diboride ($MgB_2$) shows the highest critical temperature ($T_c$ = 39 K) among the intermetallic compounds [1]. Therefore, related intermetallic compounds have attracted attention as potential new superconductors. $MgB_2$ is composed of a graphite-structured hexagonal boron (B) layer and a layer of magnesium (Mg) atoms. Since electrons are transferred from Mg to B, the formal charge of B is −1 with electronic configuration of $2s^2 2p^2$. The metallic state is derived via electron transfer from the $\sigma$ bands made up of B $p_{x,y}$ orbitals to the $\pi$ bands containing B $p_z$ orbitals [2-4]. We can find similar anionic $p$-block metal layers in other intermetallic superconductors. For example, the anionic Bi ($6s^2\ 6p^4$) square net produced by Bi $6p$-Bi $6p$ $\sigma/\pi$ bonding is responsible for the superconductivity of $R$Ni$_x$Bi$_2$ ($R$ = La, Ce, Nd and Y) compounds [5]. Moreover, the superconductive Zintl phases such as $AE$Sn$_x$ ($AE$ = Sr, Ba) have intermetallic anionic Sn-layered structures containing $\pi$ bonding systems [6, 7].

In this paper, we report the investigation of superconductivity (critical temperature $T_c$ = 0.6 K) in $Ca_2InN$ [8] containing a 2-dimensional (2D) chain array of anionic In ($5s^2 5p^{1+\delta}$). Density functional theory (DFT) calculations and X-ray photoemission spectroscopy (XPS) measurements indicate that the superconductivity arises from the anionic state of the In layer [-(In$^{0.6-}$)-] which is formed by In $5p$ orbitals with $\pi$ and $\sigma$ bonding states.

## 2. Experimental details

The polycrystalline $Ca_2InN$ was synthesized by the reaction of $Ca_2N$ and In metal. All handling was performed in an argon-filled glove box with oxygen and moisture levels below 1 ppm. The starting $Ca_2N$ powder was synthesized by solid-state reaction between $Ca_3N_2$ powder and Ca shots as described in reference [8]. A mixture of $Ca_2N$ and In was pressed into a pellet and the pellet was wrapped with molybdenum foil and sealed into a silica glass ampoule filled with argon gas at atmospheric pressure. The ampoule was then heated at 873 K for 50 h. For homogeneity, the obtained samples were ground and heated again under the same conditions. Sintered $Ca_2InN$ polycrystalline samples were tinted dark brown, and decomposed when exposed to air.

The crystalline phase and structure were identified by powder x-ray diffraction (PXRD) using Mo K$\alpha$ radiation at 300 K. The sample was ground into a powder and placed in a glass capillary ($\phi$ 0.5 mm) to reduce effects of preferential orientation of crystallites. Rietveld refinement of the PXRD patterns was performed using TOPAS software [9].

The temperature dependence of dc electrical resistivity ($\rho$) and heat-capacity ($C_p$) were measured in the temperature range of 0.4-300 K both without and with a static magnetic field of up to 2 kOe, using a physical property measurement system (PPMS) with He$^3$ refrigeration system. Electronic band structure, density of state (DOS) and Fermi surface calculations were performed using the generalized gradient approximation (GGA), Perdew-Burke-Ernzerhof (PBE) functional, and the projected augmented plane-wave (PAW) method [10, 11] implemented in the Vienna *ab* initio simulation program (VASP) code [12]. A primitive unit cell containing two chemical formula units was used, and the plane-wave basis-set cutoff was set to 900 eV. For Brillouin-zone integrations to calculate the total energy and DOS, $8 \times 8 \times 20$ Monkhorst-pack $k$ point grids were used [13]. To obtain the projected DOS, the charge density was decomposed over the atom-centered spherical harmonics; Wigner-Seitz radii of 0.20(1) nm for In, 0.16(8) nm for Ca and 0.08(3) nm for N were used for the PDOS calculation. The Fermi surface was visualized using Xcrysden software [14]. The electron localization function (ELF) was calculated to clarify the chemical bonding state by identifying whether the electrons were localized or delocalized.

The oxidation state of $Ca_2InN$ was confirmed by XPS and ultraviolet photoelectron spectra (UPS) measurements using a hemispherical analyzer [x-ray (Mg K$\alpha$ line: $h\nu = 1253.6$ eV) source and Non-monochromatic UV photons (He I line: $h\nu = 21.2$ eV) source]. To prepare a clean surface, the sample was cleaved under high vacuum. All XPS spectra were measured at room temperature and displayed as a function of the electron binding energy with respect to the Fermi level. The core level spectra were subsequently deconvoluted using a least squares Gaussian fit method.

## 3. Results and discussion

All peaks in the PXRD pattern could be indexed to an orthorhombic structure (space group: $C$mcm) with lattice parameters: $a$ = 0.352(9) nm, $b$ = 2.020(7) nm and $c$ = 0.496(9) nm (See figure S1 and table S1 of supplementary data). These values agreed well with those in the literature [15]. Figure 1 shows the crystal structure of $Ca_2InN$, made up of alternate $Ca_2N$ layer and one dimensional zigzag chains of In layers along the $b$-axis [15]. The $Ca_2N$ layer consists of edge shared $Ca_6N$ octahedra and contains two non-equivalent Ca sites respectively shared by two and four octahedra. Each $Ca_2N$ layer has separation of 0.468 nm, which is 20 % larger than that found in $Ca_2N$ (0.386 nm) [8]. The In layer embedded between the $Ca_2N$ layers forms an array of zigzag chains parallel to the $c$-axis. The In-In-In bond angle is 115.12° and In-In bond length is 0.294 nm, which is 10 % shorter than the shortest separation (0.325 nm) in metallic In. The inter-chain In-In distance (along the $a$-axis) is 0.353 nm, which is 20 % larger than the intra-chain In-In bond length.

Figure 2(a) shows the temperature dependence of electrical resistivity. Metallic behavior, with resistivity decreasing with decreasing temperature, was seen; the resistivity value at room temperature was 27 mΩ·cm. A sudden drop to zero resistivity occurred at 0.7 K; this drop was not seen when a magnetic field of $H$ = 2 kOe [inset of figure 2(a)] was applied. Figure 2(b) shows specific heat data (plotted as $C_p/T$ vs. $T^2$), below 1.2 K, collected at $H$ = 0 and 2 kOe. A jump was observed from $T$ = 0.7 K without magnetic field, but was completely suppressed when $H$ = 2 kOe. These results clearly indicate that bulk superconductivity appears from 0.7 K. The specific heat above $T = 0.7$ K when $H = 0$ Oe, i.e., in the non-superconducting state, can be fitted to the formula:

$$C_p / T = (C_e + C_l) / T = \gamma + \beta T^2 \quad (1)$$

where $C_e$ and $C_l$ denote electronic and lattice contributions to the total specific heat [16]. The obtained $\gamma$ and $\beta$ values were 4.24 mJ/mol K$^2$ and 1.39 mJ/mol K$^4$, respectively. The inset of figure 2(b) shows the electronic specific heat data (plotted as $C_e/T$ vs. $T$) for Ca$_2$InN. $T_c^{mid.} = 0.6$ K was defined as the midpoint of the specific heat jump, and the height of the heat capacity jump ($\Delta C_p^{obs.}$) was 2.85 mJ/mol K. The Debye temperature ($\Theta_D$) is estimated as 322 K by using:

$$\Theta_D = (12\pi^4 nR / 5\beta)^{1/3} \quad (2)$$

where $n$ is the number of atoms per formula unit of Ca$_2$InN and $R$ is the gas constant. To evaluate the electron-phonon coupling constant ($\lambda_{el-ph}$), we use McMillan's expression[17]:

$$T_c^{mid.} = (\Theta_D / 1.45) \, exp\{[-1.04(1 + \lambda_{el-ph})] / [\lambda_{el-ph} - \mu^*(1 + 0.62\lambda_{el-ph})]\} \quad (3)$$

where the $\mu^*$ is the retarded Coulomb repulsion parameter and usually amounts to 0.1- 0.2. Substituting 0.1 and 0.2 for $\mu^*$, we get $\lambda_{el-ph} = 0.36$ for 0.1 and $\lambda_{el-ph} = 0.53$ for 0.2, and find that Ca$_2$InN is classified as a weak-coupling BCS-type superconductor [17].

According to BCS theory for weak coupling, the superconducting volume fraction was evaluated to be 78% from

$$\Delta C_p^{obs.}(T_c) / \Delta C_p^{BCS}(T_c) \quad (4)$$

where the theoretical heat jump at $T_c$ ($\Delta C_p^{BCS}$) is predicted to 3.64 mJ/mol K from BCS theory.

$$\Delta C_p^{BCS}(T_C) / \gamma T_c = 1.43 \quad (5)$$

Figure 3(a) shows the calculated electronic band structure, with total and projected DOS

(PDOS) per formula unit (f. u.), for $Ca_2InN$. Here, we noted that the narrow entangled bands located from −3 to −1 eV with Ca $d$ character did not appear in the band structure calculated by extended Hückel method in the past report [15]. This difference is due to a flaw of the calculation method ignoring the contribution of Ca $3d$ orbitals. The four bands which cross the $E_F$ and form a DOS peak around $E_F$ (−0.9 eV < $E$ < 2.0 eV) are mainly made up of In $5p$ bands with small contributions from the atomic orbitals of Ca and N, as shown in the PDOS. The energy levels of N and Ca are far from the $E_F$, and In $5p$ bands are inserted between the N $2p$ and Ca $3d$ bands. By integrating the PDOS on the In $5p$ orbital up to $E = E_F$, we find that the valence of In in $Ca_2InN$ is $In^{0.6-}$, i.e., the electronic configuration of In may be expressed as $[Kr]4d^{10}5s^25p^{1.6}$. This calculated result indicates that the anionic indium layer includes a $p$-band, which is stabilized by the strong electron donating ability of the $[Ca_2N]^+$ cationic framework. The cationic $Ca_2N$ layer is also found in dicalcium nitride $Ca_2N$ regarded as 2D electride $[Ca_2N]^+e^-$, while the connecting pattern of $Ca_6N$ octahedron is a little different [8, 18]. The ionicity of $Ca_2N$ layer is so strong that electrons are pushed out to empty inter-layer space.

Figure 3(b) shows the Fermi surface of $Ca_2InN$. The two kinds of cylindrical sheets along Γ-Y, which are parallel to the direction of the alternate stacking of $Ca_2N$ and In layers ($b$-direction in conventional cell), are perpendicular to the In-chain array direction ($a$- and $c$-direction in conventional cell). Additionally, one kind of the cylindrical sheets around the R and S points consist of flat planes along the Γ-Z direction, which are perpendicular to the In-chain direction ($c$-axis; figure 1). This result suggests that the embedded In zigzag chain (quasi-one-dimensional) array in $Ca_2InN$ has an electronic structure with 2D character, with some flat areas in the Fermi surface because of the binding state within the In-chain array. The 3D character, which occurs along the Γ-Z direction, comes from the orbital interactions between In $5p$ and Ca $3d$, N $2p$. The electronic structure can also be seen in the calculated ELF in figure 3(c). The large ELF values around Ca and N atoms reflect that the chemical bond between Ca and N atoms is ionic. In contrast, the small ELF values spread around the In-chain direction and inter-chain direction indicate that the electrons are 2D delocalized along the In-chain array, and are independent from the $Ca_2N$ layer.

Figure 3(d) shows orbital angular momentum projected DOS on In site. The In $5p_x$, $5p_y$ and $5p_z$ orbitals contribute to the four entangled bands near $E_F$. These bands are dispersed along Γ-S, R-Z (along the $a$-direction) and S-R (along the $c$-direction). These results imply that the electronic properties of $Ca_2InN$ are attributable to the $p\sigma$ and $p\pi$ bonding complex in the $a$-, and $c$-directions (In-chain array plane) of the In $5p$ orbital, which is occupied by electrons supplied from the $[Ca_2N]^+$ cationic layer. The In-In distance along In-chain shorter than that in In metal is consistent with the participation of $p\sigma$ and $p\pi$ bonding nature to the intra In-In bonding. In other words, electrons mainly occupy the bonding molecular orbitals of –$(In)_n$– polymer and the bond length is shortened as a result. This is in contrast to the bonding state of indium metal in which each In atom is loosely bound by metallic bonding mediated by 3-dimensinally delocalized $5s$ and $5p$ electrons. Therefore, we identify that the superconductivity in $Ca_2InN$ originates from the In $5p$ state of the In-chain array. The strength of intra-chain bonding has been confirmed by means of crystal orbital overlap populations calculation in the reference [15]. Because the anionic In ($5s^2 p^{1.6}$) state induces a complex $\sigma$ and $\pi$ bonding state, and the In structure has a low dimensionality electronic structure because of the In zigzag chain array, the In $5p$ state in $Ca_2InN$ has a more narrow band dispersion than that found in either the 3D cationic In $5s$ or 3D anionic In $5p$ states. Actually, the 2D anionic In ($In^{0.6-}$) state in $Ca_2InN$ has a narrower band width (~4.1 eV) than that of the 3D anionic In ($In_5^{7-}$) band in $Sr_3In_5$ zintl phase (~5 eV) [19].

Finally, the charge state of In was examined by XPS. Figure 4 shows the In $3d$ XPS spectrum of the sample. Two peaks are identified as In $3d_{3/2}$ at 451.0 eV and In $3d_{5/2}$ at 443.5 eV. A shoulder at 444.5 eV was assigned to $In^{3+}$, originating from the surface state [20, 21]. These observed In $3d$ binding energies show a significant lower energy shift compared to those of indium oxide (444.5 ± 0.1 eV; +3 charge state) and pure indium (443.9 ± 0.1 eV; neutral state) [22-24]. This finding indicates that the charged state of In in $Ca_2InN$ is negative, which is compatible with DFT calculations ($In^{0.6-}$), and demonstrates that $Ca_2N$ electride acts as a strong electron donor.

Since the $Ca_2InN$ and In metal are weak coupling BCS type superconductor [25-26], the $T_c$ can be estimated by the equation:

$$T_c = 1.13\, \Theta_D \exp(-1/VN(E_F)) \qquad (6)$$

where $\Theta_D$, $V$ and $N(E_F)$ are Debye temperature, the paring potential of electron-electron interaction and the density of state at the $E_F$, respectively. As evidence from equation (6), $N(E_F)$ and $V$ are the dominant factors for $T_c$ determination, whereas $\Theta_D$ plays a minor role. By putting experimental values of $T_c$ and $\Theta_D$, and the calculated value of $N(E_F)$ by DFT to equation (6), the value of $V$ was estimated. Table 1 summarizes the parameters associated with superconductivity of $Ca_2InN$ and In metal. While $V$ is closein these systems, $\Theta_D$ of $Ca_2InN$ is 3 times higher than that of In metal, corresponding to strength of In-In bond. However, $N(E_F)$ of $Ca_2InN$ is 0.5 times smaller than that of In metal. This small $N(E_F)$ is the main reason for lower $T_c$ of $Ca_2InN$, though 2D structure is generally considered to be favorable to enhance $N(E_F)$. This result implies that $Ca_2N$ block layer is rather thick to get the high volume density of In chain.

4. Summary

We reported a BCS-type superconductivity with $T_c$ = 0.6 K in $Ca_2InN$ composed of 2D anionic In-chain array embedded between $Ca_2N$ layers. By assuming a BCS-type weak coupling, the superconducting volume fraction was estimated to be ~80%. From DFT calculations, we conclude that the superconductivity arises due to the anionic state of the In layer [-($In^{0.6-}$)-] which is formed by In 5p orbitals with $\pi$ and $\sigma$ bonding states. The anionic state of In was experimentally confirmed by the XPS chemical shift. The strong ionicity of $Ca_2N$ layer could be applied for stabilizing various anionic metal layer forming $p\sigma$ and $p\pi$ bonding system similar to MgB$_2$ and related compounds. If an appropriate $p$-block elemental is chosen and embedded instead of In, we expect that the resulting compound would also exhibits superconductivity.

**Acknowledgments**

This work was supported by the JSPS FIRST project and the MEXT Element Strategy Initiative project.

**Figure Captions**

**Figure 1** Crystal structure of $Ca_2InN$. Green, light blue and gold spheres represent Ca, N and In atoms, respectively. Solid and dashed dark brown lines represent the conventional and primitive unit cell. Bond angle and distances were obtained from Rietvelt analysis.

**Figure 2** Physical properties of $Ca_2InN$. (a) Temperature dependence of electrical resistivity ($\rho$). Inset shows $\rho$ - $T$ curves under magnetic fields of 0 (open circle) and 0.2 kOe (open square). (b) Low temperature heat capacity under magnetic fields of 0 (open circle) and 0.2 kOe (open square). Inset is the temperature dependence of the electronic specific heat data divided by temperature, $C_e / T$. The solid gray curve is the theoretical curve according to BCS theory [16].

**Figure 3** Calculated electronic structure of $Ca_2InN$. (a) Band structure and total density of states (DOS) and atomic projected DOS (PDOS). For the band structure, the green, blue and red circle size lines on the left-hand side of the figure denote the widths of the atomic distributions of Ca, In and N. In the DOS, the black line denotes the total DOS of $Ca_2InN$, and the green, blue, orange and red lines denote the Ca $d$, In $p$, In $s$ and N $p$, DOS, respectively. (b) Fermi surface of the first Brillouin zone. Some of the high symmetry points are shown. (c) Electron-localization function (ELF) of the total electron density on the (2 0 0), (0 0 3) plane parallel to the $b$-axis. Solid dark brown lines represent the primitive unit cell. (d) PDOS of In $5p$. The black, pink and blue line thicknesses indicate the widths of the distributions of energy of the $p_x$, $p_y$, and $p_z$ components of the In $5p$, respectively. Beside the PDOS, a schematic of the bonding state on the zigzag In chain is shown. The gold spheres represent the In atoms.

**Figure 4** X-ray photoemission spectrum of $Ca_2InN$ (In $3d$ region). Solid lines (blue and red) are the results of Gaussian deconvolution of the peak. The inset shows the He I ($h\nu$ = 21.2 eV) ultraviolet photoelectron spectrum (UPS) valence band of $Ca_2InN$.

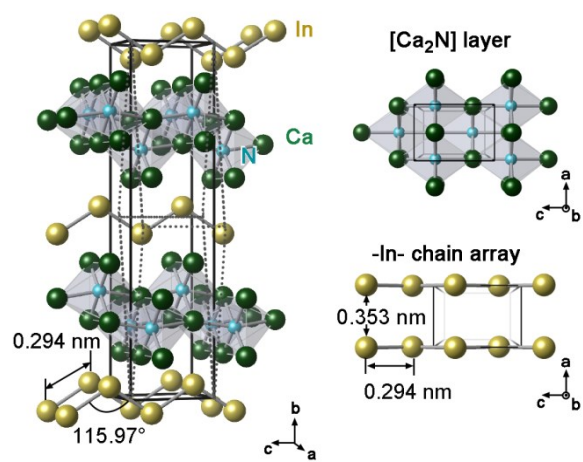

**Figure 1**

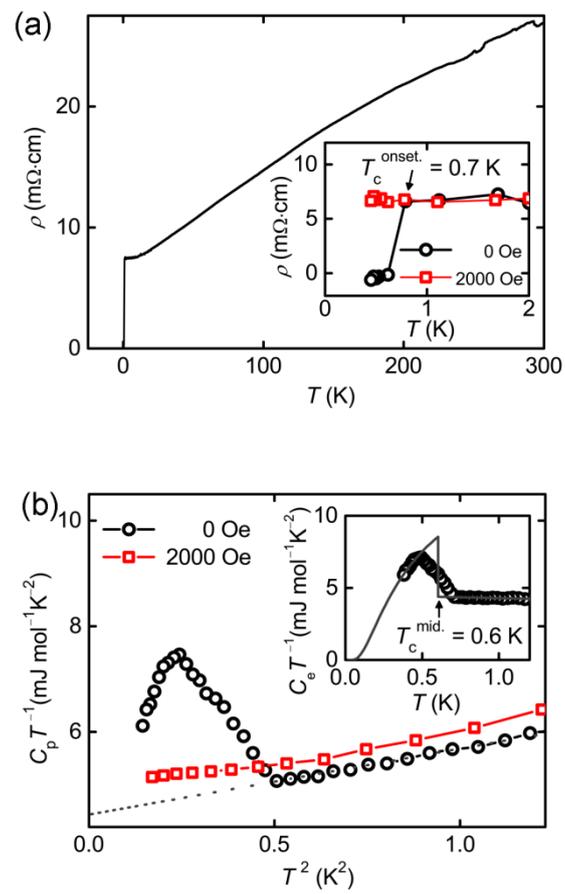

**Figure 2**

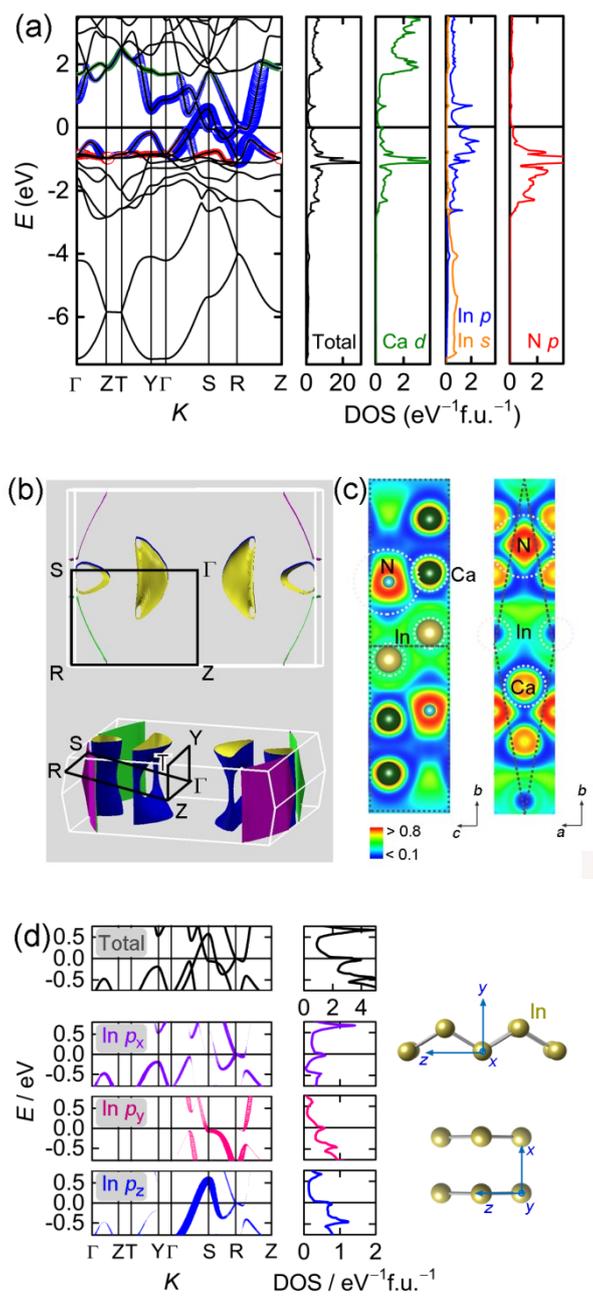

**Figure 3**

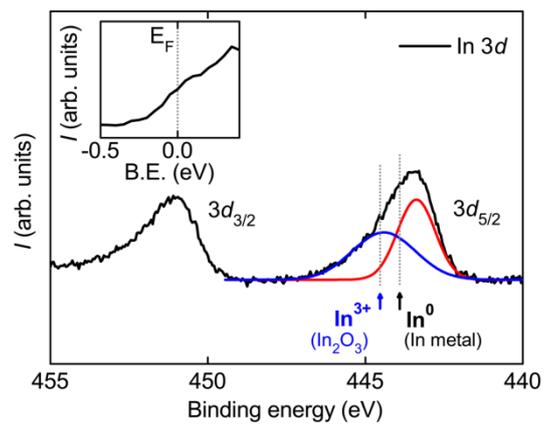

**Figure 4**

**Table 1.** Comparison of the superconducting parameters between $Ca_2InN$ and In metal.

| $Ca_2InN$ | Parameter | In metal |
|---|---|---|
| 0.6 | $T_c$ (K) | 3.41 |
| 322 | $\Theta_D$ (K) | 108 |
| 2.29 | $N(E_F)$ (eV$^{-1}\cdot$nm$^{-3}$) | 4.59 |
| 0.068 | $V$ | 0.058 |